\documentclass[aps,prb,twocolumn,groupedaddress]{revtex4}

\usepackage{graphicx}
\begin{document}


\title{Octapartite parameterization for the spin-${1\over 2}$ square
  lattice Heisenberg antiferromagnet}



\author{Bayo Lau} \author{Mona Berciu} \author{George A. Sawatzky}
\affiliation{Department of Physics and Astronomy, University of
British Columbia, Vancouver, BC, V6T 1Z1} \date{\today}


\date{\today}

\begin{abstract}
We introduce a novel parameterization of the Hilbert space of a
spin-${1\over 2}$ Heisenberg antiferromagnet using an octapartite
description of the square lattice. This provides a systematic way to
model the ground-state wavefunction within a truncated basis. To prove
its effectiveness we study systems with up to 64 spins using exact
diagonalization. At significantly reduced computational cost, we get
ground-state energies within 1\% of the best published values. Its
non-iterative nature and lack of exploitation of spatial symmetries
make this approach suitable for generalization to doped systems.
\end{abstract}


\maketitle


{\em Introduction:} The discovery of the high-$T_c$ cuprates has
emphasized the need to understand the physics of charged fermions moving in a
two-dimensional (2D) background of antiferromagnetically (AFM) coupled
spins, as represented by a doped Hubbard
model.\cite{leerev} Since an analytical solution seems unattainable,
numerical modeling of large-scale, strongly correlated doped systems
is of extreme importance and has been pursued in many different ways,
each with their own advantages and disadvantages.

For example, the powerful Monte Carlo (MC) methods give extremely
accurate information for the undoped AFM, providing the benchmark for
the ground state (GS) energies and correlation functions of systems
with up to thousands of spins.\cite{qmc0} However, they are
limited by the sign problem if additional fermions are present. This
lead to work on alternative optimized ways for treating the undoped
AFM, with the latest development in MC sampling being the use of
variational RVB-type ans\"atze.\cite{rvb0,rvb1,rvb2} In
the same context, density matrix renormalization group (DMRG) has also
progressed over the years.\cite{dmrg}

Since the interactions with the fermions are often comparable or
bigger than the AFM's energy scale, most such AFM-specialized methods
require very non-trivial modifications to adapt to doped systems, due
to technicalities of the RVB ansatz or the DMRG special boundary
conditions. In fact, recent studies of doped systems still perform
exact diagonalization (ED) of the full Hilbert space rather than
trying to take advantage of either of these schemes.\cite{ed_h,ed_tj,n40}
ED studies have the huge advantage that they provide the GS {\em
wavefunction}, from which any GS properties, including all correlation
functions, can be calculated. This would make ED the numerical
method of choice, were it not for the extreme restriction on the sizes
of doped systems that can be currently
treated.\cite{ed_tj,n40}

A very different way to approach the doped systems is to start from a
non-optimal AFM background description, like the classical N\'eel
state.\cite{ed_ising_tj} However, the strong quantum fluctuations
leading to strong deviations from a N\'eel-like background are
believed to be an essential part of the low-energy physics of the
doped systems.  It is therefore necessary to find accurate yet
efficient ways to describe the AFM background, which can also be
straightforwardly extended to the doped problem.

Here we present a novel way to model the GS wavefunction
for the undoped AFM, with a high level of accuracy that can be
systematically improved.  It is so effective that the $N=64$ wavefunction can be calculated as a numerical vector
in a systematic basis with a commodity computer. The
method is formulated in the real-space basis of the square lattice with
periodic boundary conditions, in which other interactions are
defined naturally. Neither translational nor point-group
symmetries are exploited. Thus, it can be used as an excellent
starting point for studying doped models where these symmetries
are broken, or as an efficient kernel for iterative
methods, such as renormalization and quantum cluster theory.\cite{qcd}

We stress that the immediate goal is not to compete with the MC powerhouses in
dealing with large undoped systems. Rather, our method provides a general
insight in the essence of the AFM background, which can be used
to then reduce the computational cost for larger doped systems with only
slight loss of accuracy. Availability of such efficient yet accurate
methods is invaluable in allowing extensive studies of dependence on
various parameters, to shed light on their importance and effects.

In this paper we discuss the physical insights behind our method and apply it
to the undoped AFM, where its accuracy can be easily
gauged. Application to doped models
is in progress and will be presented
elsewhere.\cite{bayonew} With the AFM exchange taken as the
energy unit, a $N$-site AFM Heisenberg model on a
square lattice reads
\begin{equation}
H_{AFM}= \sum_{\langle i,j\rangle}
\bar{S_i}\cdot\bar{S_j}\label{eq:AFM}.
\end{equation}
The dimension of the Hilbert space is $2^{N}$, and
that of the commonly used $S^z=0$ basis for the GS is
$N!/(\frac{N}{2}!)^2$. It is the shear size of the Hilbert space that
makes this problem so difficult, and stimulates new approaches.

{\em The method:} After Anderson pointed out the connection between
RVB states and the projected BCS-type wavefunction, \cite{rvb0} the RVB
framework has been implemented for an AFM torus by two approaches. The
bottom-up route consists in the explicit optimization of the bond
amplitudes.\cite{rvb1}
The top-down approach is the optimization of projected
wave functions,\cite{rvb2} (this showed
the coexistence of superconducting and AFM order parameters for
underdoped cuprates). So long as the AFM order parameter is finite,
there should be another way to describe the wavefunction without
the projecting ansatz or the micro-managing of bond amplitudes.

One way to specify a singlet with finite AFM order is to divide
the bi-partite lattice into sublattices A and B, and perform
angular momentum addition between their states. If
these states  are classified by their total and
$z$-projection quantum numbers,
${S_{A/B}}\in[0,\frac{N}{4}],m_{{A/B}}\in[-{S_{A/B}},{S_{A/B}}]$, we
can build $S=0$ singlet states:
\begin{equation}
|0,0\rangle=\sum
\frac{(-1)^{{S_A}-m_{A}}}{\sqrt{2{S_A}+1}}\delta_{{S_A},{S_B}}
\delta_{m_{A},-m_{B}}
|{S_A},m_{A},{S_B},m_{B}\rangle\label{eq:bi_basis}
\end{equation}
The Hilbert space contains a huge number of such singlets. Since the
AFM ground-state is also a singlet,\cite{marshall} if we label all
possible singlets of Eq. (\ref{eq:bi_basis}) with an index $\alpha$,
we can express the GS wave function as their linear combination
$|GS\rangle=\sum_\alpha c_\alpha |0,0\rangle_\alpha$. We seek such an
orthonormal basis which is also computationally efficient.

The staggered magnetization is a well-studied observable of the
ground-state wave function. For a bi-partite lattice in
a staggered magnetic field, this quantity is formally defined as the
difference between $\langle m_A\rangle$ and $\langle m_B\rangle$ in the
zero-field limit. In the absence of a staggered field, this quantity has
alternative definitions, such as:
\begin{equation}
\hat{m^2}= \frac{1}{N^2}\left(\sum_r (-1)^{|r|} \hat{S}_{r}\right)^2 =
\frac{\left(\hat{S}_{A}- \hat{S}_{B}\right)^2}{N^2},
\label{eq:sm}
\end{equation}
where $\hat{S}_{A/B}$ are the total sublattice  spin
operators.  In the AFM GS, $m=\langle \hat{m^2}\rangle^{1\over 2}$
has been extrapolated to $\sim0.3$ as
$N\rightarrow\infty$, and  increases like
$1/\sqrt{N}$ to  $\sim0.45$ for $N=32$.\cite{afmrev}  In
terms of the total  spin $\hat{S}=\hat{S}_{A}+
\hat{S}_{B}$, we have:
\begin{equation}
\hat{m^2}=\frac{1}{N^2}(2\hat{S}_{A}^2+2\hat{S}_{B}^2-\hat{S}^2)
\label{eq:sm_dev}
\end{equation}
The ground state is a singlet, $S=0$. It follows that the wave
function must allocate significant weight to sectors with high values
of $S_A=S_B$ in order to yield such large values of $m$. In fact, for
the RHS of Eq.~{\ref{eq:sm_dev}} to be larger than the expected value,
the sublattice spins ${S_{A/B}}$ have to be within $\frac{N}{16}$ of
their maximum values.
Our computational basis takes advantage of this
observation. There are many ways to add spins quantum
mechanically, but not all  are viable because the truncation error
needs to be bounded systematically and the  computational
effort must be small. There are two extremes: if the original
single-site basis of
Eq.~({\ref{eq:AFM}}) is used, no basis
transformation is needed but enforcing truncation based on
Eq.~({\ref{eq:sm_dev}}) is costly. However, if a random basis
tabulated according to values of $S_A$ is used, transforming
Eq.~({\ref{eq:AFM}}) into the new basis would be costly due to the
many Clebsch-Gordon series needed. We propose a parametrization which
is a good compromise between these two extremes.

\begin{figure}
\includegraphics[width=0.7\columnwidth]{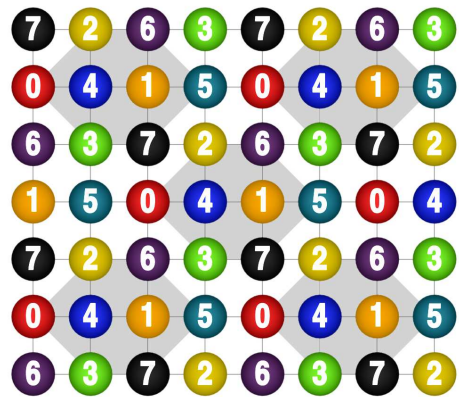}
\caption{\label{fig:octa} (color online) Square lattice divided in
  octads (shaded areas). Sites positioned similarly inside octads have
  the same label. Each is surrounded by
  neighbors with different labels.}
\end{figure}

Starting from the case where all $\frac{N}{2}$ sublattice spins
add to the maximum  of $\frac{N}{4}$, if we want to include
states with spin down to
$\frac{3N}{16}=\frac{N}{4}-\frac{N}{16}$, the basis must allow many
new  configurations, including ones where $\frac{N}{8}$ spins have
a total spin of 0 while the other $\frac{3N}{8}$ spins have a total
spin of $\frac{3N}{16}$. This suggests that groups of $\frac{N}{8}$ or
fewer spins must be allowed to take {\em all} possible spin quantum
numbers. Since a larger group would
overshoot the sublattice spin below the $\frac{3N}{16}$ "threshold",
while a smaller group would introduce extra "enforcement" costs as
discussed above, we divide the full lattice into groups of 8 sites, see
Fig.~\ref{fig:octa}. The resulting {\em octads} repeat periodically
with translational vectors $2a(1,\pm 1)$. Each spin is
identified by the octad it belongs to, and by its position inside
the octad.  A sublattice is composed from 4 groups of spins indexed
with the same label, {\em e.g.} $\hat{S}_{A}= \hat{S}_{0}+
\hat{S}_{1}+\hat{S}_{2}+\hat{S}_{3}$.  With this
arrangement, each spin interacts with spins from all the 4 groups
of the other sublattice, {\em e.g.} a group 0 spin is always
to the west/east/south/north of a spin in group 4/5/6/7. This
partition is the minimum division that permits such a description and
is a fundamental building block for the wave function in the
model. Even though we arrived at it by looking to optimize the
computation, our partition has been used in other contexts.\cite{ft}

We now straightforwardly generalize Eq. (\ref{eq:bi_basis}) and
write singlets of the total lattice as:
\begin{equation}
|0,0,f(\{{S_i}\})\rangle= \sum
c_{\{S_i,m_i\}}\prod_{i=0}^{7}|{S_i},m_i\rangle\label{eq:basis}
\end{equation}
where $S_i, m_i$ are the quantum numbers for $\hat{S}_{i}$, {\em i.e.}
the total spin for group $i=0,7$. Here, $c_{\{S_i,m_i\}}$ is the
product of the appropriate Clebsch-Gordon coefficients for the 8 pairs
of quantum numbers $({S_i},m_i)$. The $f(\{{S_i}\})$ index on the LHS
keeps track of the many ways in which these spins can be added to form
a singlet.

If all these singlets are kept into the computational basis, the GS
calculation is exact. The total spin of each group takes values $S_i
\in [0, {N\over 16}]$, and there is no
reason to restrict them. However, our previous
discussion suggests that while all possible $S_i$ values must be
allowed, the singlets with highest GS weight are those whose
total sublattice spin is within ${N\over 16}$ of the maximum value.

We therefore parameterize the singlet Hilbert space in terms of a
completeness parameter, $C_s\in[0,1]$, and include in the singlet basis
only states for which ${S_{A/B}}\geq \frac{N}{4} (1-C_s)$. For
$C_s=1$, the calculation is thus exact.  For $C_s=\frac{1}{4}$, this means
that the maximum number of anti-aligned spins in the sublattice is
$\frac{N}{8}$. Based on the discussion for the staggered
magnetization, we  expect this to already be a good
variational basis for the GS. For any $C_s\ne 1$, this
formulation allows {\em all possible values} of ${S_i}$ for each
$i=0,7$, but restricts the ways in which they can combine to give the total
sublattice spin. The number of discarded states is combinatorially
large due to the large degeneracy of states with low sublattice spins.

Eq.~(\ref{eq:basis}) is invariant under spin rotations, thus the
anti-alignment is enforced here in terms of angular momentum
addition, very different from the classical N\'eel
picture. This approach also maintains the full translational symmetry
of the Hamiltonian. If the RVB bond-amplitude optimization is a
bottom-up way of adding AFM order to a singlet, \cite{rvb1} our method
is a compatible top-down approach without the
need of projecting an ansatz.

\begin{table}[b]
\caption{\label{tab:size} Size of the $C_s=\frac{1}{4}$ subspace as
compared to that of the commonly used basis.}
\begin{ruledtabular}
\begin{tabular}{c|c c c c}
N & $C_s=\frac{1}{4}$ & $S=0$& $S_z=0$ & Full \\ \hline 16 & 50 & 1430
&12870&$2^{16}$\\ 32 & 11041 & 35357670&601080390&$2^{32}$ \\ 64 &
$9.8\times10^{8}$ & $5.55\times10^{16}$&$1.83\times10^{18}$&$2^{64}$\\
\end{tabular}
\end{ruledtabular}
\end{table}

The cost of this truncation scheme is the one-time computation of
matrix elements between these basis states. This is acceptable
because conventional diagonalization schemes are limited by storage,
not processor speed, and this data is reusable for other
computations that involve the AFM. The octapartite
division of Fig.~\ref{fig:octa} is valuable for
greatly minimizing the number of finite matrix elements of the
Hamiltonian in the basis states of
Eq.~(\ref{eq:basis}). The two-body interaction can be computed
efficiently because only 2 out of 8 kets in Eq.~(\ref{eq:basis}) are
changed and the delta functions in the Clebsch-Gordan coefficients of
Eq.~(\ref{eq:basis}) allow data to be aligned in a computationally
efficient way. The overall matrix can be constructed column-by-column
and term-by-term and is
extremely sparse. The practical implication  is that
the grand problem is decomposed into many small independent
computations which can be massively parallelized.  Table~\ref{tab:size}
compares the $C_s=\frac{1}{4}$ basis size to those of other bases commonly used by
non-iterative methods. At $C_s=\frac{1}{4}$, the $N=32$ system was solved
within seconds using a generic Lanczos routine. Even for the
$9.8\times10^8$-size basis for $N=64$, the GS vector can be
calculated with a power method using less than 15 matrix-vector
products, each taking less than 30 minutes with a single cpu thread on
a xeon workstation. Another advantage is the absence
of the gapless $S=1$ magnon excitations from the variational space,
which allows the matrix-vector products to converge rapidly.
Since these excitations
are Goldstone modes, the required $S=1$ basis to describe them has roughly the
same size as the $S=0$ basis required for the GS, and can be generated
 by appropriately changing the Clebsch-Gordon coefficients
in Eq. (\ref{eq:bi_basis}).  The modeling of {\em collective} magnon
excitations in the presence of carriers is model-dependent
and will be discused elsewhere.\cite{bayonew}

\begin{figure}[b]
\includegraphics[width=1\columnwidth]{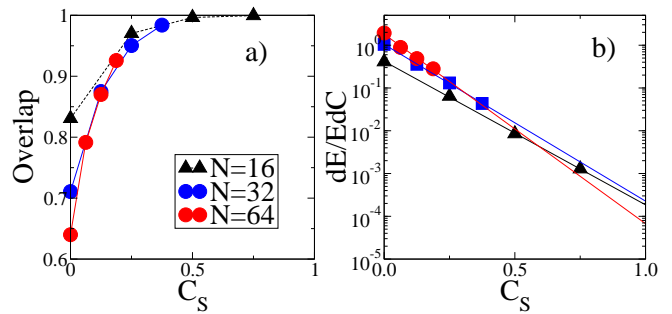}
\caption{\label{fig:v} (color online) (a) Overlap $|\langle GS,
C_s|GS, C_s+{2\over N}\rangle|^2$ between ground-state wave functions
corresponding to adjacent values of $C_s$ for $N=16, 32, 64$. Note
that for $N=64$, the overlap is already 95\% for $C_s={1\over 4}$; (b)
Fractional change $\frac{1}{E_{GS}}\frac{dE_{GS}}{dC_s}$ in
ground-state energy corresponding to different $C_s$ cutoffs. The
lines are linear fits to the data.  }
\end{figure}

{\em Results:} We have performed computations for the full basis for
$N=16$ as a benchmark for larger $N$ values. For $N=32$ (64) we
computed up to $C_s=\frac{1}{2}$ $(C_s=\frac{1}{4})$. While the
computations with $C_s< 1$ are variational, the goal here is to
demonstrate that $C_s\sim\frac{1}{4}$ is a good rule-of-thumb value to
capture well the AFM ground-state.

The stability of this formulation is demonstrated in
Fig.~\ref{fig:v}(a) which shows the overlap between the GS
wave functions for consecutive values of $C_s$. The deviation
decreases exponentially with increasing  $C_s$, and for
$C_s=\frac{1}{4}$ the overlap is $\approx$ 95\%. We conclude
that the  GS vector is already pointed in the correct direction
even for small $C_s$, and subsequent increments of $C_s$  merely
result in minor improvements. This is reinforced by the convergence of
the GS energy demonstrated in Fig.~\ref{fig:v}(b), which
supports the scaling law $\frac{dE_{GS}}{dC_s} \sim e^{-\alpha
C_s}$. The error decays exponentially, with a rate that increases
with $N$. The $N=32, 64$ lines cross at $C_s\sim
\frac{1}{4}$ where the fractional error is $\sim 10\%$. This
exponential efficiency is amplified by the fact that a linear decrease
in $C_s$ leads to a combinatorial decrease of the basis size because
of the numerous low-spin combinations  removed from the basis.

\begin{figure}[t]
\includegraphics[width=1.0\columnwidth]{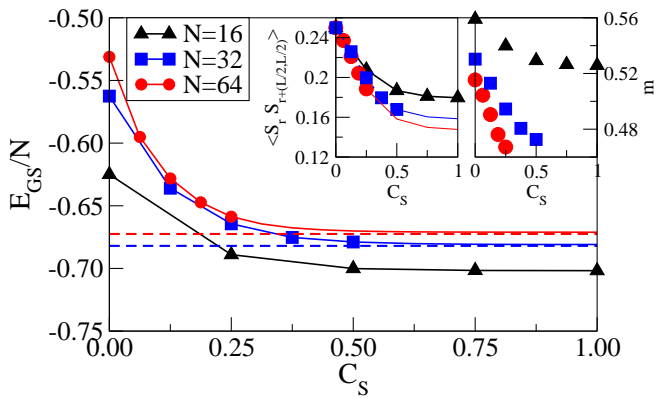}
\caption{\label{fig:energy} (color online) GS energy
  computed at specific $C_s$ values (symbols) and the extrapolation to
  $C_s=1$ (solid lines) for $N=16,32,64$. The horizontal dashed lines
  are the lowest values from Ref. \onlinecite{afmrev}. Inset:  expectation
  values of the non-commuting operators $\sqrt{\langle m^2\rangle}$ and
  $\langle\overline{S}_r\cdot\overline{S}_{r+(\frac{L}{2},\frac{L}{2})}\rangle$.}
\end{figure}

These GS energies are not as
accurate as for established iterative methods, however a single-pass
computation at $C_s=1/4$ already has less than $2\%$
error. Using a linear fit of $\frac{dy}{dx}\sim e^{-x}$, an estimate of
$E_{GS}$ at $C_s=1$ is obtained by simple integration
(see Fig.~\ref{fig:energy}).  From the  $C_s=1$ estimates and the
$N^{-\frac{3}{2}}$
scaling, we extrapolate $\lim_{N\rightarrow \infty}{E_{GS}\over N}=
-0.6671$, within $0.5\%$ of the best published value of
-0.6692,\cite{afmrev} even though it is
achieved at a significantly reduced computational cost.

We have thus far demonstrated the exponential convergence of the GS
eigenvalue and eigenvector with increasing $C_s$, which validates our
proposed variational basis.  Because MC studies cannot produce an
explicit wavefunction, they often gauge accuracy via spin-spin
correlation operators which do not commute with the Hamiltonian.
The insets of Fig.~\ref{fig:energy} show expectation values of
two such operators (the staggered
magnetization of Eq.~{\ref{eq:sm_dev}} and the spin-spin correlation
at maximum distance) evaluated from our eigenvector for a given value
of $C_s$. Although these operators are non-commuting,
their matrix representation is diagonal in our basis and
therefore trivial to compute.  The convergence of these quantities is
essentially linear for $C_s<\frac{1}{4}$.  For greater values of
$C_s$,
$\frac{d}{dC_s}\langle\overline{S}_r\cdot
\overline{S}_{r+(\frac{L}{2},\frac{L}{2})}\rangle$
is suppressed faster than $\sim e^{-C_s}$  so the post-threshold convergence is
  even better than for the GS energy.
On the other hand, because $m$ is
exploited to discard low-spin parts of the Hilbert space, its
approximated value is always higher than the true value; convergence
is only asymtotical for $C_s>\frac{1}{4}$.
Thus, the computational discount for the GS
wavefunction at a small $C_s$ is achieved at the cost of lowered accuracy
for one of many definitions of staggered magnetization,  which
does not commute with ${\cal H}_{AFM}$.

{\em Conclusions:} We introduced an octa-partition of the square
lattice, and used it to build an orthonormal singlet basis for
calculating the AFM GS.  We used $C_s$ to systematically parameterize
the resulting Hilbert space, and proposed apriori arguments to
postulate that $C_s\sim\frac{1}{4}$ is the minimal value that captures
the GS accurately. The resulting basis is combinatorially smaller than
in any other schemes, apriorily known, and computationally very
efficient. We demonstrated the stability of the formulation in
Fig.~\ref{fig:v}(a) and identified the sources of error. Using a
single thread on a commodity computer, single-pass matrix computations
are feasible for systems with up to $N=64$ spins, significantly
exceeding the current full ED record of $N=40$.\cite{n40}

While other established methods can yield smaller errors for the GS
for larger $N$ values, they cannot be easily generalized. Our
formulation is non-iterative and is defined in the real space
coordinate of a torus without exploiting spatial symmetries. This
allows for an economic generalization of this approach into iterative
procedures.

 One concern about Hilbert space truncation is its applicability to
 complicated models. Our formulation by itself does not involve MC
 sampling and so it is not subject to the sign problem. One
 outstanding issue in cuprates is the explanation and measurement of the
 photoemission spectrum at low doping and low temperature.\cite{ronning}
 Information about the spectrum is contained in the Krylov space of
 the model Hamiltonian acted on the AFM GS;\cite{damascelli} thus, a
 compact description of the AFM GS in a systematic orthonormal space
 can indeed narrow the space spanned by calculations. Because the
 total spin of the AFM plus carriers is conserved while AFM order is
 expected to persist for low doping at low temperature,  our method
 can be generalized to reduce the
 Hilbert space of the doped systems. The details are model-dependent
 and outside the scope of this paper; this work is progress.

{\em Acknowledgement:} This work was supported by NSERC, CIfAR and CFI.

\end{document}